# Market Crashes as Critical Phenomena?
## Explanation, Idealization, and Universality in Econophysics

Jennifer Jhun
Department of Philosophy
Lake Forest College

Patricia Palacios
Munich Center for Mathematical Philosophy
Ludwig-Maximilians Universität München

James Owen Weatherall
Department of Logic and Philosophy of Science
University of California, Irvine

Abstract:

We study the Johansen-Ledoit-Sornette (JLS) model of financial market crashes (Johansen, Ledoit, and Sornette [2000,]. "Crashes as Critical Points." *Int. J. Theor. Appl. Finan* **3**(2) 219-255). On our view, the JLS model is a curious case from the perspective of the recent philosophy of science literature, as it is naturally construed as a "minimal model" in the sense of Batterman and Rice (Batterman and Rice [2014] "Minimal Model Explanations." *Phil. Sci.* **81**(3): 349–376. ) that nonetheless provides a causal explanation of market crashes, in the sense of Woodward's interventionist account of causation (Woodward [2003]. *Making Things Happen*. Oxford: Oxford University Press).

1. Introduction

Mainstream economic models of financial markets have long been criticized on the grounds that they fail to accurately account for the frequency of extreme events, including market crashes. Mandelbrot and Hudson (2004) put the point starkly in their discussion of the August 1998 crash: "By the conventional wisdom, August 1998 simply should never have happened. The standard theories estimate the odds of that final, August 31, collapse, at one in 20 million, an event that, if you traded daily for nearly 100,000 years, you would not expect to see even once. The odds of getting three such declines in the same month were even more minute: about one in 500 billion" (p. 4). Similar critiques have been mounted in connection with the October 1987 "Black Monday" crash and the October 1997 "mini-crash", as well as other large drawdowns over the last thirty years. By the lights of ordinary economic reasoning, such events simply should not occur.

Motivated in part by the October 1987 crash, and in part by new interdisciplinary initiatives in nonlinear dynamics during the late 1980s, the last thirty years has seen an upswell in alternative approaches to economic modeling, many of which have been inspired by analogies with statistical physics. Work in this tradition has come to be known as *econophysics*, a term coined by H. Eugene Stanley in 1996.[1] One goal of econophysics has been to develop new financial models that can accurately describe, and perhaps even predict, extreme events such as financial crises.

Despite the apparent empirical successes of some models in econophysics, the field has not been widely embraced by economists. The few who have engaged have been strongly critical. For example, Lo and Mueller (2010) have argued that econophysics is doomed because "human behavior is not nearly as stable and predictable as physical phenomena" (1), and thus the strategies available in physics are not at all suitable for dealing with economic phenomena.[2] Similarly, Gallegati et al. (2006) express skepticism about the assumption – essential for econophysics – that "universal empirical regularities" (1) of a sort amenable to predictive mathematical modeling are to be found outside of narrowly isolated subfields in economics.[3] Worse, they argue, the models of statistical physics that econophysics draws on tend to have conserved quantities, such as energy, that are exchanged in interactions; in economics, by contrast, it is precisely the quantities of exchange, such as money, that are *not* conserved. Thus, one of the central "idealizing assumptions" of (some) econophysics models seems to be strongly contradicted by basic facts of economic life. Finally, they accuse econophysicists of using unrigorous statistical methods to identify patterns that are not really there.

In this paper, we make the case that econophysics deserves more credit than these critiques give it, as an enterprise that is at least sometimes successful in its main goals of predicting and explaining economic phenomena of certain kinds. Our strategy will not be to address the general criticisms just described head-on, and we do not mean to argue that all models from econophysics, or even most or many models, are successful. Instead, we will focus on just one model that, we will argue, has two features of interest: it (1) draws on a significant analogy with statistical physics, in a way that goes beyond standard modeling methods in economics; and (2) has real explanatory, and possibly predictive, power. Our principal goal is to elaborate and defend how we take the model to work, including where and how the analogy with statistical physics enters, and to articulate what sorts of novel insights into market behavior we believe it offers. In this sense, we take the model we consider as "proof of concept", while simultaneously providing a case-study for the sorts of explanatory goals that arise in econophysics.

---

[1] For more on the relationship between physics, finance, and econophysics, see Weatherall (2013); for further technical details and overviews of recent work, see Mantegna and Stanley (1999), McCauley (2004), and Cottrell et al. (2009). There is also a small literature in philosophy of science dealing with econophysics, including Rickles (2007) and Thebault, Bradley, and Reutlinger (forthcoming).

[2] Despite the prevalence of this sort of criticism, it is far from clear that physics is more guilty of oversimplification than economics when it is applied to economic facts.

[3] Gallegati et al (2006) has sparked a small debate, with responses by McCauley (2006)and Rosser Jr. (2008) among others. It is worth noting that financial markets, which are the focus of the present paper, are one of the areas of economic activity that Gallegati et al. seem to think are amenable to the methods of econophysics, and so it is not clear that the model we consider here is touched by these criticisms.

The model we consider is the Johansen-Ledoit-Sornette (JLS) model of "critical" market crashes (Johansen, Ledoit, and Sornette 2000), which uses methods from the theory of critical phase transitions in physics to provide a predictive framework for financial market crashes.[4] This model is of particular interest because it aims both to predict and describe market-level phenomena – crashes – and to provide microscopic foundations that explain how that behavior can result from interactions between individual agents. More specifically, in addition to its predictive role, the JLS model aims to explain two "stylized facts" associated market crashes.[5] The first is the fact that stock market returns seem to exhibit power law behavior in the vicinity of a crash, and the second is so-called *volatility clustering*, which is the fact that market returns seem to exhibit dramatic, oscillating behavior before crashes, with large changes followed by other large changes.[6]

The plan of the paper is as follows. In section 2, we will present some (limited) background on mainstream modeling in financial economics that will help place the JLS model in context.[7] In section 3, we will introduce the model itself, focusing on the role the analogy with critical phase transitions plays in the model. Then, in section 4 we will argue against one tempting way of understanding how the model works, and instead defend a somewhat different understanding. On the view we will defend, the principal achievements of the model are to explain why crashes occur endogenously in markets and to provide a possibly predictive signature for impending crashes.

Central to our argument in section 4 will be the observation that although the analogy with critical phase transitions is crucial in motivating and developing the model, in the end the analogy is only partial. In particular, although the model fruitfully draws on the renormalization group theory of critical exponents, financial crashes do not seem to constitute a universality class in the strict sense that one encounters in that area of physics. Nonetheless, we argue, there is a weaker sense in which crashes exhibit universal features. This weaker notion of universality allows one to draw novel inferences about the microscopic mechanisms that might underlie crashes. Since the model helps make salient the possible microscopic mechanisms that could explain the occurrence of a crash, we claim that the model provides an explanation of crashes that is both causal (in the sense of Woodward (2003)) and reductive.

---

[4] For more on this model and related ideas, see especially Sornette (2003) and references therein. The model has since been elaborated and widely discussed; a recent discussion of criticisms and replies is given by Sornette et al. (2013).

[5] These stylized facts are often treated as qualitative laws or as descriptions of lawlike behavior, capturing "set[s] of properties, common across many instruments, markets, and time periods" (Cont 2001, 223).

[6] This has also been noted by Mandelbrot (1963, 418).

[7] For more on how the JLS model fits into mainstream financial modeling, see Sornette (2003); for background on mathematical methods in finance more generally, see for instance, Joshi (2008).

In section 5 of the paper, we will explore how the argument just sketched relates to recent debates in philosophy of science concerning explanatory uses of idealized models. We will argue that the JLS model is naturally understood as a "minimal model" in the sense of Batterman and Rice (2014) (see also Batterman 2002, 2005, 2009). Nonetheless, we claim, (apparently) contra Batterman and Rice, that it provides both a causal and reductive explanation of market crashes. As we will argue, this shows that the same mathematical methods may be used for multiple explanatory purposes, and that to understand explanatory strategies in the context even of minimal models, one needs to pay careful attention to the salient why questions.[8]

We conclude with some remarks about possible policy consequences. In particular, we argue that our interpretation of the JLS model as one that yields causal explanations suggests methods by which policymakers could intervene on the economy in order to prevent crashes or to halt the spread of one. The JLS model, we argue, may be used as a diagnostic tool, allowing economists and regulators to formulate new measures or to assess the performance of ones are already in place.

## 2. Some Financial and Economic Background to the JLS Model

Although the JLS model draws extensively on methods and ideas from the theory of critical phenomena in physics, it also builds on a long, mainstream tradition of market modeling in financial economics. Moreover, Sornette and collaborators emphasize this continuity with early work in financial modeling. In the course of analyzing work in econophysics, it seems particularly important to be clear about just where this work diverges from more traditional modeling. And so in this section we will provide some minimal background on methods and ideas from financial modeling that the JLS model builds on.

The JLS model may be broadly located in a tradition of modeling markets as stochastic processes. This tradition originated with groundbreaking work in 1900 by French mathematician Louis Bachelier, who first proposed treating price changes as a random walk and built an options model on this basis (Bachelier and Samuelson 2011). Bachelier's work went largely unnoticed, however, until re-discovered by J. L. Savage and Paul Samuelson in the early 1950s. Independently, in 1959 a physicist named M.F.M. Osborne proposed modeling market returns as undergoing Brownian motion (Osborne 1959). Osborne provided his own empirical support for this model, though it was largely consistent with earlier empirical work on market time series by the Cowles Commission (1933) and by Kendall (1953).

---

[8] As we hope will be clear in what follows, we do *not* mean to disagree with Batterman (2000; 2002) and others, such as Reutlinger (2014), who have argued that explanations of why universal phenomena occur that draw on RG methods are generally non-causal. Instead, we mean to argue that RG methods may be used for multiple purposes, and that in the present case, the salient explanations have a different character than in the case of statistical physics. The explanandum is not the existence of universality, and the explanation is causal.

Later, Samuelson (1965) and Fama (1965, 1970) explicitly connected the random-walk hypothesis to the *efficient markets hypothesis* (EMH).[9] The EMH is the claim that markets are informationally efficient and asset prices reflect (all) available information. The EMH is consistent with, and indeed implies, market randomness. This is because if markets are assumed to assimilate information efficiently, then any information available to market participants at a given time will already be factored into the price at that time.[10] Thus only (unaccounted for) news, which is random, changes prices, meaning that changes in stock option prices themselves must be random. Persistent exceptions to this rule, it is argued, are impossible, since if traders were to observe a pattern in asset price time series that could be exploited, they will exploit it, which would tend to wash out the pattern.

More formally, in efficient markets prices follow a martingale process, which is a general stochastic process where the conditional expectation of the next value, given past history and current value, is precisely the current value. That is,

$$E(p_{t+1} - p_t | \Omega_t) = 0, \text{ where } \Omega_t = (p_1, p_2, \ldots p_t), \text{ the history up till time t.}$$

Here $E(p_{t+1} - p_t | \Omega_t) = 0$ is the expectation value of the change in price in a given time-step. Thus, for an asset that pays no dividends, one should expect the future price to hover around the current value, all other things being equal.

$$E_t[p(t')] = p(t), \text{ for all } t' > t$$

In other words, we could say that the prices of stocks do not depart from their *fundamental* or *intrinsic* value in a way that an investor could systematically predict or exploit to make a profit in the long run. In this sense, the EMH implies that the market will behave unpredictably.

The market models just described have some well-known limitations. For instance, if returns are modeled as a random walk, as Osborne and others proposed, one would generally expect returns to be normally distributed. In fact, however, market returns tend to be "fat-tailed".[11] This means that we see extreme events more often than one would expect if returns were normally distributed. In addition, treating markets as a martingale process leaves out a number features that appear to be good indicators of crises, such as volatility clustering (where large changes in price are followed by further large changes in price).

---

[9] The EMH has been a topic of considerable controversy. For instance, Shiller (1984) has argued that the argument behind the EMH is invalid. The main worry is that current models neglect (i) agent psychology and (ii) interactions amongst agents as key causal and explanatory features of asset price variations. Once these factors are considered, it seems markets may well be random irrespective of how efficiently markets process information or how accurately prices reflect fundamental values. Meanwhile, as Ball (2009) and others have argued, over-reliance on the assumption of efficiency may affect how market participants synthesize information regarding possible asset bubbles. But we will not weigh in on such controversies; our purpose here is not to endorse the EMH, but rather to describe the context of the JLS model and to emphasize its continuity with mainstream economic modeling methods.

[10] Note that this argument appears to suppose that news that will positively affect price is equally likely as news that will negatively affect price. But if there were any information available that would indicate that positive (resp. negative) news was more likely, then that fact alone would count as tradeable information that would affect price.

That said, neither the martingale condition nor the EMH is in and of itself inconsistent with fat-tailed distributions or with large asset price changes. Indeed, there is a tradition in economics of modeling *rational bubbles*, which are deviations from fundamental values that are compatible with the martingale condition and the EMH (Blanchard 1979; Blanchard and Watson 1982; Allen and Gorton 1993; Santos and Woodford 1997; Sornette and Malevergne 2001) The idea is that under some circumstances markets enter a "speculative regime" in which it is rational to hold onto an asset in anticipation of growing future returns, even though one believes that the current price is not the fundamental price. Here, markets may still be understood to be processing information efficiently – and thus the EMH may be taken to hold – since the endogenous facts about the speculative regime are themselves information bearing on future prices. In this regime, an asset's value grows indefinitely, which itself is not realistic but may be a suitable modeling assumption if persistent increase in value is anticipated over the timescale of interest.

Still, rational bubbles models of the sort just described provide no insight into the circumstances under which the speculative regime ends and markets crash. The JLS model is intended to extend rational bubbles models in order to explain and predict market crashes in the speculative regime. The basic proposal is that financial bubbles and subsequent crashes are much like the development of sudden, spontaneous, and drastic behavior in physical systems such as magnets. Like earlier rational bubbles models, the JLS model treats bubbles and crashes without rejecting the EMH. Instead, as we will see in the next section, it attempts to reconcile the EMH with a story about the behavior of interacting traders.

## 3. The JLS model

Important stock market crashes of the twentieth century, including the US crashes of 1929 and 1987 and the Hong-Kong crash of 1997, have been the result of the action of a large group of traders placing sell orders simultaneously. Curiously, this synchronized "herding" behavior seems to arise endogenously, rather than from outside instruction or the influence of communication media. Traders, who are geographically apart and generally disagree with each other, seem to organize *themselves* to place the same order at the same time. The JLS model concerns the character and dynamics of this self organization between traders.[12]

In physics, critical phase transitions constitute an important class of phenomena that likewise exhibit "self organization". A paradigm example of these kinds of transitions is the paramagnetic-ferromagnetic transition in magnetic materials. In this transition, a large group of spins that are generally pointing in different directions align themselves in the same direction simultaneously, so that the system undergoes spontaneous magnetization. This suggests a potentially useful analogy between critical phase transitions and stock market crashes.

---

[11] See, for instance, Mandlebrot (1963) and Cont (2001).
[12] Note that we mean "self-organization" in the informal sense of coordinated action between agents without any apparent external mechanism. We do not intend to invoke any specific theories of self-organization or self-organized criticality.

Motivated by this analogy, Johansen et al (2000) propose a model (henceforth the JLS model) that elaborates on the rational bubbles models noted in the previous section and other work in econophysics (e.g. Sornette et al (1996) and Sornette and Johansen (1997)). The main hypothesis underlying this model is that market crashes may be understood as a "critical phenomenon" strongly analogous to critical phase transitions. This hypothesis is made precise by postulating a correspondence between the quantities that are used to describe financial crashes and the physical quantities that describe critical phase transitions. This correspondence then allows one to draw inferences concerning various quantities of interest, including the probability of a crash occurring under various circumstances.

In more detail, on the JLS model a stock market crash occurs when the system transitions between two phases: a phase *prior to the crash* and a phase *after the crash*. This transition point is analogous to the critical point for physical systems, and in the present context corresponds to the time at which a stock market crash is most likely to occur. In this model, there are two quantities that are relevant for capturing this behavior of interest. The first is known as the *hazard rate*, $h(t)$. The hazard rate measures the instantaneous rate of change of the probability of the event occurring at time $t$, given that it has not yet occurred by $t$. The larger the hazard rate, the more rapidly the probability of an impending crash is increasing, given that the crash has not yet occurred.[13] It may be thought of as the instantaneous rate at which crashes should be expected to occur, if only crashes were repeatable. The second quantity is the price of some asset as a function of time, $p(t)$. These two quantities determine the dynamic equation that will be used to predict future crashes and provide a framework for the underlying microfoundational story.

The model begins with a general form for the price dynamics for a time prior to a crash. These dynamics are given by:

$$\log \frac{p(t)}{p(t_0)} = k \int_{t_0}^{t} h(\tau) d\tau, \qquad (1)$$

where $p(t_0)$ is the price at some initial time $t_0$, $p(t)$ is the price at a subsequent time $t$, $k$ is a constant, and $h(t)$ is the hazard rate. Note that the hazard rate determines the price.[14] This means: the higher the hazard rate, the faster the price of an asset will rise. In other words, the more risky the asset is, the more the trader expects to receive in the future as compensation for taking on that risk.

---

[13] More precisely, if $F(t)$ is the cumulative distribution function of a crash occurring at or before time $t$, then $h(t) = F'(t)/(1 - F(t))$, where $F'(t) = dF/dt$ is the probability density function. Conversely, one can define a cumulative probability function from a hazard rate by integrating both sides of this equation with respect to $t$. See, for instance, Cleves et al (2004, Ch. 2) for further details on interpreting hazard rates.

Note that these dynamics are consistent with the standard financial modeling assumptions described above. In particular, in the special case where the hazard rate vanishes, the expected change in price over any given time interval vanishes, just as one would expect from the martingale condition discussed in Section 2 for a stock that does not pay dividends. Following JLS, we call this the "fundamental regime". When the hazard rate is positive, meanwhile – the so-called "bubble regime" – one expects price to increase exponentially over time. In this regime, the increase in price is driven up by the accumulated risk involved in holding the asset during a period in which a crash is deemed possible. Investors are willing to pay ever higher prices on the grounds that they expect price to continue to increase without bound, as long as a crash does not occur.

In this general form, these dynamics do not give an account of stylized facts such as the power law behavior we observe in financial time series, nor do they tell us anything about the microscopic mechanism underlying the occurrence of a crash. It is to get these further results that one introduces the qualitative analogy to critical phase transitions. (Up to this point, no such analogy has been invoked.)

To begin, we suppose that markets consist of populations of two types of traders, which JLS call "rational" and "noise" traders. (It is not essential that these populations be distinct; particular traders may sometimes be noise traders and sometimes rational traders.) The rational traders are assumed to trade on the basis of market fundamentals; noise traders, meanwhile, are assumed to base their decisions on trends, imitate others around them, etc. rather than investigating market fundamentals (Kyle 1985; Black 1986).

The model then assumes that traders are situated in a lattice network, analogous to the lattice of the Ising model, the most important model in the study of phase transitions, including the paramagnetic- ferromagnetic transition mentioned above. (Note, however, that the specific lattice structure will turn out to be distinct from the Ising model.) Agents in this network may be in one of two possible states: a "buy" state or a "sell" state, just as spins in an Ising model may be either "up" or "down." Also like in an Ising model, agents are assumed to imitate their nearest neighbors, so that if a given agent is in a different state from the average of her neighbors, there will be a non-zero probability that the agent will change states. A crash on this model is understood as a moment in which a large group of traders are suddenly in the "sell" state.[15] Therefore, in this model a crash is caused (at the microscopic level) by self-reinforcing imitative behavior between traders. This behavior is analogous to a phase transition, during which a large number of nodes in the Ising model adopt the same state.

---

[14] It is tempting to interpret the right hand side of Eq. (1) as representing the probability of a crash occurring during the period from $t_0$ to $t$, but this would be incorrect: the integral of $h(t)dt$ does not yield a probability. (For instance, it may exceed 1.) Instead, this quantity should be understood as a measure of accumulated risk, in the sense that it represents the total number of times you should have experienced a crash during this period, supposing the crash were repeatable. Once again, see Cleves et al (2004, Ch. 2).

In statistical mechanics, the quantity that best describes the tendency of particles to imitate one another is the *susceptibility* of the system. In the ferromagnetic-paramagnetic transition mentioned above, this quantity corresponds to the magnetic susceptibility $\chi$, which is governed by the following power law near the transition point:

$$\chi \approx A |T - T_C|^{-\gamma} \qquad (2)$$

where A is a positive constant, $T_C$ corresponds to the critical temperature, and $\gamma$ is known as the critical exponent. Informally, the susceptibility of the system characterizes the tendency of the system's average magnetization (which is related with the number of spins in the same state) to change due to the influence of a small external field. One consequence of the power law is that at the critical point, $T=T_C$, $\chi$ diverges. The divergence of the magnetic susceptibility implies the divergence of the correlation length, a quantity that measures the average distance over which particles in the system interact. It is due to the divergence of the correlation length at the critical point that distant particles are likely to be mostly in the same state at the same time.[16]

The JLS model posits that the hazard rate $h(t)$ has the same general form as the magnetic susceptibility

$$h(t) \approx B |t - t_c|^{-\alpha} \qquad (3)$$

where $t_c$ is the most probable time for the crash, $B$ is a positive constant, and $\alpha$ is a critical exponent that is assumed to have values between zero and one. Note that attributing this form to the hazard rate is really an ansatz: no claim has been made to have *derived* this power law behavior from any microscopic model (or family of models). Instead, we have made two independent assumptions: the first is that traders may be modeled as agents on a lattice with two states, without specifying any details of the lattice or interactions between agents; and the second is that the hazard rate has a particular form analogous to the magnetic susceptibility. The idea that the hazard rate should be analogous to susceptibility is motivated by the idea that a crash should correspond to large correlation lengths, but this does not fix the form of Eq. (3).

---

[15] Sornette (2003) also considers the possibilities of "anti-crashes", wherein a large number of traders suddenly transition to "buy" states; these are taken to be the ends of "anti-bubble" regimes. However, it is important to note that neither Sornette (2003) nor Johansen (2000) explain the fact that crashes are generally caused by "sell" states instead of "buy" states.

[16] For more details on the logic of critical phenomena in physics, see Wilson and Kogut (1974), Goldenfeld (1992), Cardy (1996), Fisher (1998), Kadanoff (2000), Sornette (2006), and Zinn-Justin (2007); for a more philosophical take, see Batterman (2002) and Butterfield and Bouatta (2015).

The final ingredient of the model is phenomenological. Observing the stylized fact that prices exhibit accelerating oscillations in the lead up to a crash, one infers that the critical exponent $\alpha$ in Eq. (3) is complex.[17] A complex critical exponent modifies the power law to include periodic oscillations in time known as log-periodic oscillations.[18] JLS argue that, to leading order in a Fourier expansion near $t_c$, the general solution for $h(t)$ is given by:

$$h(t) \approx B_0 |t_c - t|^{-\alpha'} + B_1 |t_c - t|^{-\alpha'} \cos[\alpha'' \log|t_c - t| + \psi] \qquad (4)$$

where $B_0$, $B_1$, and $\psi$ are real constants, $\alpha'$ is the real part of $\alpha$, and $\alpha''$ is the imaginary part of $\alpha$.

Having identified this form for the hazard rate, one then plugs $h(t)$ from Eq. 4 back into the general dynamic equation (1) to obtain an expression that describes the behavior of price as a function of time given this hazard rate, to obtain:

$$\log[p(t)] \approx \log[p_c] - \frac{k}{\beta} \{ B_0 (t_c - t)^\beta + B_1 (t_c - t)^\beta \cos[\omega \log(t_c - t) + \phi] \}, \qquad (5)$$

where $\beta = 1 - \alpha' \in (0,1)$, $p_c = p(t_c)$ is the price at the critical time, and $\phi$ is another constant.

Eq. 5 succeeds in capturing the stylized facts observed in the occurrence of extreme events, including volatility clustering and accelerating oscillations (Yalamova and McKelvey 2011). Moreover, as we will elaborate below, it provides an explanation of these observed phenomena – and indeed, of crashes themselves – that appeals to the existence of self-reinforcing imitative behavior between traders. Finally, the model aims to be predictive by providing the tools to anticipate the occurrence of crashes that arise due to endogenous herding behavior, such as panics, by describing a specific form of accelerating oscillations – namely log periodic oscillations – that provide a signature of approaching criticality.

---

[17] The argument here is subtle. JLS first present their model generically, without making any assumptions about the details of the network. They then observe that if the network has certain features – in particular, if it is *hierarchical* in a sense to be explained in section 4.2 – then it will exhibit complex critical exponents, and hence log-periodic oscillations near criticality. They give some plausibility argument for considering hierarchical lattices, but leave the actual lattice structure open until they consider historical data – at which point they conclude that, given the presence of oscillations, the network must be approximately hierarchical and the critical exponents must be complex. It is in this sense that introducing complex critical exponents is "phenomenological". One can also run the argument in the other direction, however, and argue that on the basis of a plausible assumption concerning the hierarchical nature of trader networks, the critical exponents should be expected to be complex; at times, Sornette and collaborators appear to prefer this version of the argument.
[18] An early discussion of log-periodicity and self-similarity is given by Barenblatt and Zel'dovich (1972). Extensive work on the existence of complex critical exponents with log-periodic oscillations has been carried out by Sornette and his collaborators (eg. Sornette 1998; Arnéodo, Muzy, and Sornette 1998; Gluzman and Sornette 2002; Zhou et al. 2005; Sornette 2006).

Note that although volatility clustering and accelerating oscillations are taken as stylized facts that are "inputs" for the model that are used to establish that the complex exponent in Eq. (3) is complex, the specific form of Eq. (5) should be taken as an output of the model. As such, it can be back-tested to provide empirical support for the model as a whole, and specifically for the claim that crashes may be understood as critical phenomena. The results of these tests have been reported in several places (Sornette, Johansen, and Bouchaud 1996; Sornette and Johansen 1997; Johansen, Ledoit, and Sornette 2000; Sornette 2003; Graf v. Bothmer and Meister 2003; Calvet and Fisher 2008). The model has also been used to provide real-time predictions of market crashes (Sornette et al. 2015). Perhaps most remarkable is the crash of 1987, where the log-periodic oscillations are visible even to the naked eye (Johansen, Ledoit, and Sornette 2000).

## 4. The Logic of the JLS model

The JLS model, and the analogy between crashes and critical phenomena on which it is based, are highly suggestive. However, one needs to be careful about the limits of the analogy. [19] As we will presently argue, even if one accepts the arguments given in the previous section, the logic of the model is importantly different from that of models from statistical physics on which it is based. First, we will argue that unlike critical phase transitions, "critical" market crashes do not form a universality class in the sense of renormalization group (RG) physics. It follows that explanatory strategies familiar from applications of the RG in physics do not carry over directly to this model. We will then present a different analysis of the logic of the JLS model, emphasizing what sort of explanations we think the model can provide. We will conclude by observing that although the mathematical methods used in the JLS model are similar to those from physics the role that these methods play in application are different.

### 4.1 Do Market Crashes Constitute a Universality Class?

To evaluate the analogy between market crashes in the JLS model and critical phenomena in physics, we will begin by describing the situation in physics in some further detail. As noted above, when a system undergoes a critical phase transition, some important physical quantities diverge. For instance, in the ferromagnetic-paramagnetic transition described in section 3, the divergent quantities are the magnetic susceptibility, the specific heat, and the correlation length. The divergence of the correlation length implies that all spins are correlated at the transition point regardless of the distance between them. That is, the measuring distance unit is no longer important. When this happens, the system is said to be *scale invariant*.

---

[19] There are various criticisms of the JLS model that also stress the disanalogies between the JLS model of financial crises and critical phase transitions. For example, Ilinski (1999) casts doubt on a main component of the JLS model: crashes are principally caused by imitative dynamics between individual traders. He objects that different market participants may act over different time horizons (e.g. minutes for speculators, years for managers), so that the instantaneous long-range interactions between traders postulated by the JLS model are implausible. We will not engage with this criticism or others; instead, we want to see how far the analogy goes if we assume that the model *is* well-motivated and well-supported empirically.

Scale invariance is consistent with the observation of power law behavior of physical quantities near a critical point. The exponents appearing in these power laws – called *critical exponents* – were originally determined experimentally. Surprisingly, radically different systems, such as fluids and ferromagnets, were found to have exactly the same values for their critical exponents. This was particularly striking because the exponents were deemed *anomalous*, which is to say that they were not whole numbers or simple fractions. Systems having the same values of their critical exponents are said to belong to the same *universality class*.[20] One of the great achievements in the theory of phase transitions was the development of RG methods to explain how this universal behavior comes about – i.e., to explain why apparently different systems have the same scaling behavior near criticality.

RG methods consist, roughly, in a set of transformations by which one replaces a set of variables by another set of – generally coarse-grained – variables without changing the essential physical properties of a system. The (infinite) iteration of these transformations in a space of Hamiltonians enables one to find so-called *fixed points* of the transformation, which are Hamiltonians that represent the (coarse-grained) dynamics of a system near a transition point.[21] This procedure is taken to explain universality, as it has been shown that systems in the same universality class flow to the same fixed points, and thus the systems in a given universality class should be expected to have the same dynamical properties near the transition point. The existence of non-trivial fixed points is generally taken to show that a system's microscopic details are irrelevant to its behavior near criticality. In addition, RG methods provide an argument for the use of highly idealized models in the explanation of radically different systems. For instance, by showing that both ferromagnets and fluids are in the same universality class as the Ising model, RG methods justify the use of the Ising model for the study of both systems.

---

[20] As will become clear in what follows, by "universality class" we mean the basin of attraction of a given non-trivial fixed point under some RG flow. In cases of critical phase transitions, these correspond to systems with the same critical exponent near the transition point, though RG methods may be applied more generally. Batterman and Rice (2014) suggest a still-broader definition of "universality class" that applies to systems outside of physics where the RG does not apply; as we will see below, market crashes will turn out to form a universality class in this more general sense, but one needs to be careful about the role that the RG plays in the argument for this.

[21] Note that our description of RG methods here follows the "field space" approach, in the sense of Franklin (2017).

Thus, in physics, the logic of universality arguments goes as follows. One begins with the empirical observation that certain systems exhibit the same behavior – i.e., have the same critical exponents – near criticality.  One then shows that those that systems flow to the same fixed point by iterated application of an RG transformation, thus explaining their observed similarity by establishing that, at a certain level of coarse-graining, these systems have the same dynamical properties.  In other words, the thing one is ultimately trying to explain is why a range of apparently different systems are all saliently the same, and the explanation proceeds by showing that the microscopic details of the systems do not matter to the phenomenon in question.[22]

Is the same reasoning applied in the JLS model?   It would seem not.  In particular, the first step does not work.  While data-fitting supports the idea that the relationship between price returns and hazard can be captured via a power-law (e.g. Johansen, Sornette, and Ledoit 1999), analysis of past crashes does *not* support the hypothesis that crashes constitute a universality class in the sense of all corresponding to the same non-trivial fixed point of some RG flow.  This is because crashes do not all exhibit the same critical exponent. Via curve-fitting, Graf v. Bothmer and Meister (2003) show that in 88 years of Dow-Jones-Data there actually are no characteristic peaks in the critical exponent $\beta$ of equation 5. Although JLS showed that the exponent of the crash in 1987 and the crash in 1997 differ by less 5%, Sornette et al. (1996) show that the value of that exponent differs substantially from other important crashes such as the crash in 1929. The fact that that there is no characteristic peak in the exponent $\beta$ has the following consequence. Stock market crashes are neither in the same universality class as the Ising model (or any previously solved model) nor do they constitute a universality class themselves.

One might think, as Sornette and collaborators themselves seem argue to in at least some places, that the fact that crashes do not constitute a universality class entirely undermines the analogy between crashes and critical phenomena.

> "If we believe that large crashes can be described as critical points and hence have
> the same background, then $\beta$, $\omega$ and $\Delta t$ should have values which are comparable."
> (Johansen, Ledoit, and Sornette 2000, p. 2397)

As we will argue below, however, we do not think that the failure of crashes to constitute a universality class is a major problem for the model.[23]  But it does mean that the logic of the model, and the sorts of explanations we can expect from it, are importantly different from in physics. If we cannot expect crashes to constitute a universality class, then the RG story cannot be applied either for the calculation of critical exponents or for the explanation of the universal behavior observed in crashes (or not observed, as it happens). In other words, if there is universal behavior in stock market crashes, this is not the kind of universal behavior that can be explained via RG methods alone.[24]

---

[22] Note that it is not essential, here, to begin with an empirical observation – though that is what happened in the physics of phase transitions.  In principle, one can demonstrate that two systems are in the same universality class and thereby predict their behavior near critical points.

[23] In addition to what we argue in what follows, Sornette (personal correspondence) points out that market crashes should be understood as dynamical (i.e., non-equilibrium) phase transitions, wherein a parameter diverges at a critical time.  In these systems, one generally finds universality to be much weaker than in equilibrium systems.

## 4.2 On the Explanatory Character of the JLS model

We saw above that the JLS model apparently does not work by establishing that market crashes form a universality class. This means that one cannot apply the same reasoning as in physics to argue that large-scale market behavior near transition points (i.e., crashes) is independent of the microscopic details of market dynamics. It thus seems that insofar as the JLS model is successful, it must function differently. In this section we will develop a positive account of the logic of the JLS model, describing what we take the model to explain and how. We will argue that the JLS model relies on a subtle interplay between microscopic and macroscopic considerations, by which known mathematical facts familiar from statistical physics are used, in conjunction with empirical considerations, to draw inferences in both directions.

Recall that, whereas the arguments from statistical physics sketched above began with a brute empirical claim—many systems appear to have the same critical exponents—the JLS model began with two separate ingredients. The first, Eq. (1), was taken from mainstream economics—or at least, from the theory of rational bubbles. The second, Eq. (3), was a bare *ansatz*, inspired by statistical physics but in no sense justified by it. In other words, one begins by considering what market dynamics *would* look like *if* the hazard rate were governed by a power law near crashes, similarly to how the magnetic susceptibility behaves. These two ingredients, along with the further specification that the exponent in Eq. (3) is complex, then lead to Eq. (5), concerning the logarithm of market prices near a crash. It is this equation that is the principal predictive output of the model, and also the means by which the model is both calibrated and tested against historical data.

But this is not the whole model. To motivate the *ansatz* that the hazard rate satisfies Eq. (3) near crashes, JLS include a third ingredient, which is that microscopic market dynamics may be modeled as a network of agents, interacting with their nearest neighbors via imitation, and that the hazard rate may be interpreted, much like the magnetic susceptibility, as a measure of the characteristic distance scale of correlations between agents. The proposal that market participants form *some* sort of network of influence is taken as *prima facie* plausible, and no particular evidence is offered for it; at this stage, no claims are made about the details of the network structure. Drawing on known results from statistical physics, JLS then observe that networks of this sort are very often associated with power laws near criticality for the parameter that is now being interpreted as hazard rate, thereby linking Eq. (3) with a class of microscopic models.

---

[24] Note however that this does not mean that RG methods cannot be applied at all in the context of the JLS model. Zhou and Sornette (2003), for instance, use renormalization group methods to obtain an extension of equation (5) that gives an account of larger time scales. Moreover, as we will see, RG methods will reappear in our analysis below, although they will play a different role than in statistical physics.

One then argues that insofar as Eq. (5) is successful, this relationship between network models and power laws lends further plausibility to treating market microdynamics with a network model of this sort, and also that spontaneous herding, which now is understood to correspond to long-distance correlations in a network, explains endogenous market crashes. In particular, the divergence of the hazard rate at the critical point implies the divergence of the correlation length, i.e. the range of interaction between traders.

As we noted above, if the correlation length in a network model of this sort diverges, the system becomes scale invariant. It is under these circumstances that the system is successfully described by power laws. Scale invariance means that, near the critical point, market dynamics are self-similar across scales. In other words, as traders imitate their neighbors, they aggregate into clusters (e.g. companies) that act as individual traders imitating their neighbor companies, and so on, to higher and higher scales. This imitation procedure across different scales accounts for how information propagates so quickly before a crash: "...critical self-similarity is why local imitation cascades through the scales into global coordination" (p. 32).

But now, recall that the critical exponents in the JLS model were determined to be complex, and that the associated power laws exhibited log-periodic oscillations. Not all network models lead to log-periodic power laws (LPPLs); they typically arise (only) when the underlying network model exhibits *discrete* scale invariance. Discrete scale invariance means that the system is scale invariant only under special discrete magnification factors; this, in turn, implies that the system and the underlying physical mechanisms have characteristic length scales. As Sornette (1998) points out, this provides important constraints on the underlying dynamics. In particular, it suggests that traders are arranged on a *hierarchical* lattice, which is a lattice in which, by virtue of the network structure, some nodes (traders) have greater influence than others (still via nearest-neighbor interaction).[25] Examples of hierarchical networks such as the Bethe lattice, a fractal tree, or hierarchical diamond lattice. These hierarchical networks tell us not only how information propagates through scales but also how information propagates within the same scale. In figure 1, for instance, one can see that in the Bethe lattice information that starts by one agent propagates within the same scale faster than exponentially.

---

[25] For a general overview of hierarchical lattices and a discussion of their properties, see, for instance, Griffiths and Kaufman (1982) and Melrose (1983).

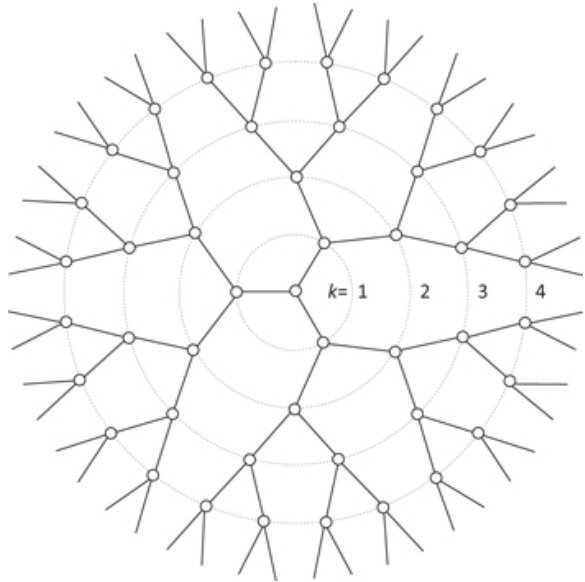

**Figure 1:** Illustration of a Bethe Lattice, one of the possible network structures underlying the occurrence of a crash according to the JLS model. The point in the center of the figure represents a trader who is source of opinion. The first ring represents the neighbors, who tend to imitate the opinion of the trader at the center. The second ring represents their neighbors, who are indirectly influenced by the opinion of the first trader, and so on. This aims to illustrate how imitation could possibly propagate resulting into global coordination.

Sornette has argued that it is plausible to model the propagation of information in social structures using hierarchical lattices, and also that there is independent empirical support for doing so (Sornette 2003, Ch. 4). But it is not claimed – nor is it necessary to claim – that under general circumstances, the network of traders is fractal, or that it corresponds to some exact hierarchical lattice. Instead, what is claimed is that under general circumstances, the network of traders lies in the basin of attraction of a hierarchical model under RG transformations, so that its critical behavior is the same as that of a hierarchical network, i.e., so that near a crash markets exhibit LPPLs. In other words, interactions between traders must be "approximately" hierarchical, in the sense of lying in the same universality class as *some* hierarchical network (with imitative dynamics). It is here that RG methods enter explicitly into the JLS model. One might think of the role played by RG methods here as establishing that crashes form a universality class in a more general sense than that discussed above, namely by showing how a wide range of systems flow to fixed points characterized by hierarchical networks of one sort or another. (We will return to this idea below.)

We claim that it is the inference from observed LPPLs to discrete scale invariance of an underlying network structure (or, more generally, from power laws of any kind to scale invariance) that forms the explanatory core of the JLS model. In more detail, what we find here is an explanation of (endogenous) market crashes as arising from the structure of the network of traders at the time the crash occurs. Markets crash in the absence of any external, coordinating event because the network of traders can spontaneously evolve into states that are (discretely) scale invariant, i.e., which have long correlation lengths, so that small, essentially arbitrary perturbations, can propagate rapidly across scales.[26]

Perhaps surprisingly given the literature on universality and explanation, this explanation, as we understand it, is *causal*, in the sense of Woodward's interventionist account of causation (Woodward 2003).[27] On Woodward's account, causes are variables that one could intervene on in order to reliably influence a system. More precisely, one says that A causes B if (given some background conditions) there is a conditional of the form "if A, then (likely) B", where A can be understood as a single variable that one could, in principle, manipulate. On this account of causation, a relationship such as the one between LPPLs, discrete scale invariance, and transitions, which holds across a range of different condition, can serve as a guide to identifying causal relations. As Woodward puts it, "When a relationship is invariant under at least some interventions…it is potentially usable in the sense that…*if* an intervention on X were to occur, this would be a way of manipulating or controlling the value of Y" (16).

We take it that the moral of the JLS model in its most general form is as follows: *if* the network of agents participating in a market approaches a (discretely) scale-invariant state, as signaled by the appearance of LPPLs in price, *then* (it is likely that) a crash will occur. In other words, the model says that crashes occur in many different systems precisely when their (coarse-grained) dynamics become approximately discretely scale invariant. And so, it is the emergence of discrete scale invariance (or, perhaps, scale invariance more generally) that should be identified as the proper cause of the crash.

---

[26] Note that, while he continues to argue that DSI and LPPLs are important features of crashes that signal the end of bubble regimes, in more recent work Sornette has suggested that both of these may be secondary, with the fundamental signature of a crash instead being positive feedbacks, leading to power law singular behavior (of some sort or other) (Sornette and Cauwels 2015b, 2015a; Leiss, Nax, and Sornette 2015). These arguments seem to us to move beyond the JLS model as we have presented it, though we take it they are broadly compatible with the picture we sketch here of the sort of explanation these models seek to give. In particular, on this alternative view it would be the inference from LPPLs to positive feedback loops that forms the explanatory core of the model. We are grateful to Didier Sornette for drawing our attention to these more recent arguments.

[27] Sornette also speaks of this explanation as "causal": for instance, when he writes "…the market anticipates the crash in a subtle self-organized and cooperative fashion, hence releasing precursory "fingerprints" observable in the stock market prices…. we propose that the underlying cause of the crash must be searched years before it in the progressive accelerating ascent of the market price, reflecting an increasing build-up of the market cooperativity" (Sornette 2003 p. 279). As we noted in footnote 8, we do not take the claim that this explanation is causal to be in conflict with the views defended by Batterman (2000; 2002), Reutlinger (2014), or others. The claim is *not* that there is an explanation of universality in this model that is causal. Rather, the claim is that the explanation of a given crash, or even crashes in general, is causal, because the JLS model identifies how to intervene to produce a crash, or to prevent one – namely, by changing network structure.

On this view, it is the state of the network as a whole that should be understood as the cause of the crash. But one might worry that this is not an "event" or "variable" of the sort that one can intervene on. We believe it is. First, observe that on Woodward's account, it need not be possible to *actually* manipulate the variable; it need only be the case that one could imagine, within the model, changing just this feature. And indeed, in the present case, one certainly can change the state of the network so that it is no longer scale invariant (discretely or otherwise), and in doing so, one *ipso facto* moves away from the transition point. This is precisely what is needed.

More can be said on this point, however. As we will explore in the final section of the paper, we believe there are mechanisms by which an agent – say, a regulatory body – can *in fact* intervene on the network structure of market participants in order to disrupt scale invariance. If this is possible, then the conditional above not only bears a clear causal interpretation, but in fact has policy implications regarding how to deal with an impending market crash. Before turning to this point, however, we will consider how the analysis of both the logic of the JLS model and its explanatory properties that we have just provided bears on recent debates concerning explanation and universality in the philosophy of science literature.

**5. Infinite idealizations, universality, and explanation in the JLS model**

In the last section, we argued that the JLS model, though bearing important relationships to models of phase transitions in physics, relied on an argument that was importantly different, both in the sense of "universality" at play and in how inferences are drawn about the micro- and macrodynamics of markets. We also presented a positive account of both the logic of the model and the character of the explanation it offers of market crashes. As we argued, this explanation is best construed as *causal*, in the interventionist sense of Woodward (2003).

We made these arguments largely independently of the recent literature on the character of explanations in statistical physics that make use of the methods the JLS model borrows. There was good reason for this: our main contention above was that the logic of the JLS model is different from that of the models of phase transitions on which it is based. That said, there are some features of the JLS model that make it salient from the perspective of recent debates on explanation in philosophy of science. In particular, the JLS model is arguably a *minimal model* in the sense of Batterman and Rice (2014).[28] A minimal model, according to Batterman and Rice, is one that "…is used to explain patterns of macroscopic behavior across systems that are heterogeneous at small scales" (p. 349). More importantly, minimal models are "thoroughgoing caricatures of real systems" whose explanatory power does not depend on their "representational accuracy" (p. 350). Instead, the key feature of a minimal model is that it allows us to say why many different systems turn out to be saliently similar, despite their significant differences at a microscopic level.

---

[28] See Lange (2015) for a different critique of Batterman and Rice (2014) than we give here. Lange argues that Batterman and Rice cannot sustain the distinction they draw between their account and "common feature" accounts such as Weisberg's (discussed below). We take it that one *can* sustain a distinction between different explanatory goals, one of which might well be to explain why many different systems should be expected to be saliently similar to some highly idealized model, and we think that Batterman and Rice do an adequate job of explaining both how that explanatory goal can be met, and why the strategies for meeting it do not look like they are appealing to common features of a model and a target system. That said, as we will argue, in some cases a single model, including the JLS model, can be used to achieve more than one explanatory goal.

The model of critical phase transitions discussed above is a paradigm example of a minimal model in the Batterman and Rice sense. There, the goal is to explain why many different systems have the same behavior near transition points, and moreover, to show why highly idealized models, such as the Ising model, capture the essential behavior of all of these different systems. The RG played an essential role in this story. But Batterman and Rice are clear that it is not only models that use the RG in this way that are to count as minimal models: they also describe an example from biology – the Fisher sex ratio model – and argue that it is a minimal model as well. The essential feature in both cases is that one has a universality class, in the general sense of a collection of models that are all similar in some salient way, and an explanation of why all of the systems in question fall into that universality class.

We argued above that even though the RG plays a different role in the JLS model than in models of critical phase transitions, there is still a sense in which market crashes form a universality class, according to the JLS model. This universality class does not correspond to the basin of attraction of a single non-trivial fixed point under iterated applications of an RG transformation. Instead, it is a collection of systems that are all saliently similar, in the sense that they exhibit LPPLs.

Still, one can explain why a wide range of systems exhibit this same universal behavior: they all exhibit discrete scale invariance near their transition points. Moreover, RG methods play an important role in this argument. Although RG transformations do not take all of the relevant similar systems to the *same* non-trivial fixed point, they do take such systems to non-trivial fixed points with complex critical exponents, and thus LPPLs. So in this sense, the RG establishes the universality class in the salient (generalized) sense. Finally, although one cannot show that there is some idealized model that has the same critical exponent as every market crash – since not all market crashes have the same critical exponent! – one can show that there are highly idealized models, each exhibiting discrete scale invariance near transition points, that give rise to LPPLs near their transition points. It is on these grounds that we take the JLS model to be a minimal model in the Batterman-Rice sense.

The JLS model also has another feature that, though not part of the official definition of minimal models, seems characteristic of them (Batterman 2005, 2009): the JLS model relies on an *infinite idealization*. (This provides one sense in which the model "caricatures" real markets.) That is, the JLS model assumes that the network of market participants includes infinitely many agents. Moreover, this feature is *necessary* for the model as we have described it, and it is assumed in all versions of the model we know of in the literature. The reason it is necessary is that scale invariance, including discrete scale invariance, means that some property of the model must hold – i.e., be "invariant" – at *all* scales, no matter how large. Thus only an infinite model may be truly scale invariant. Likewise, only an infinite model can exhibit the sort of infinite correlation lengths that we identify with a transition point.[29]

These features of the JLS model, and especially the role that the infinite idealization plays in establishing scale invariance near the critical point, are common across applications of the RG methods. And Batterman puts considerable weight on the infinities that arise in models that use these methods: rather than anomalies to be avoided or removed, they are sources of important information.

> I'm suggesting that an important lesson from the renormalization group successes is that we rethink the use of models in physics. If we include mathematical features as essential parts of physical modeling then we will see that blowups or singularities are often sources of information. (Batterman 2009, p. 11)

It seems that something similar is going on in the JLS model: there, too, one encounters not only infinite systems, but also divergent quantities – including both the hazard rate and the correlation length between traders. And it is these blowups that signal that a crash is impending. This singular behavior is at the very core of the model.

So it seems that the JLS model has the hallmarks of a minimal model. But if so, there is a tension between what we say above and Batterman and Rice's account of how minimal models explain. In particular, Batterman and Rice emphasize that the sorts of explanations they consider are *non-causal* and *non-reductive*.[30] Moreover, they argue minimal models are not *representational,* in the sense that their success does not depend on "some kind of accurate mirroring, or mapping, or representation relation between model and target" (351). On our view, however, the JLS model *does* provide a causal explanation; moreover, this explanation is arguably both reductive and representational.

We have already seen the sense in which the JLS model provides a causal explanation: it may be understood to yield a conditional statement, the antecedent of which is a variable on which one can, in principle, intervene. Thus, on an interventionist account of causal explanation, the model appears to allow us to say that it is (discrete) scale invariance that causes market crashes—or, to put it in more evocative terms, it is herding at all scales that causes market crashes.

Some readers will balk at this claim: after all, as just noted, only infinite systems can be truly scale invariant, and realistic markets are not infinite. So, in what sense could a feature that no actual market could have cause a behavior that realistic markets exhibit? Or to put it another way, how could actual market crashes be caused by scale invariance? The answer, as we see it, is that the JLS model explains crashes by showing that in some networks, correlation lengths can become long, relative to the overall size of the network, and that when this happens, crashes become likely. It is the infinite idealization that allows one to precisely characterize the relationship between long correlation lengths, scale invariance, and crashes, and it is not clear that one could establish this relationship as neatly in a finite system as one can in the infinite system. But what the infinite system is ultimately telling us is something about the causal relationship between correlations between traders and market-wide crashes.[31]

---

[29] This is not to say that the model could not be reconfigured as one that is invariant across some scales, but not under arbitrary scale transformations. In other words, we do not mean to deny what is sometimes known as "Earman's principle", that idealized models can only be explanatory if one can imagine removing the idealization and still being able to explain the same phenomenon (Earman 2004; J. Butterfield 2011). But doing so would require substantial changes in the analysis, and would effectively produce a different model from the one under consideration. Our interest is in the explanatory role of the infinite idealization in the present version of the model.
[30] See also Morrison (2006) for a related point.

We should emphasize that, although we take this explanation to be causal, it is only on a particular account of causation (i.e., the Woodward (2003) account). Of course, there are many other analyses of causation on which this may well not be a causal explanation (Salmon 1984; Strevens 2008). More importantly, we do not claim that crashes are being explained, here, by appeal to particular details concerning interactions between individual agents. In this sense, it is not a "causal-mechanical" or "mechanistic" explanation (Craver 2006; Kaplan 2011; Kaplan and Craver 2011). Indeed, the model is not committed to any particular network model at the microscale, just a class of models that exhibit discrete scale invariance. Sornette puts the point as follows.

> It turns out that there is not a unique cause but several mechanisms may lead to DSI. Since DSI is a partial breaking of a continuous symmetry, this is hardly surprising as there are many ways to break down a symmetry. We describe the mechanisms that have been studied and are still under investigation. The list of mechanisms is by no mean exhaustive and other mechanisms may exist. (Sornette 1998, p. 247)

Thus, the model does not even include a specific account of how agents interact with one another. It is rather a generic feature of a range of possible networks that plays the causal role.

This last point is also closely related to the senses in which we take the JLS model to be reductive and representational. The antecedent of the conditional described above refers to the micro-constituents of the market. It is in this sense that we take the explanation to be reductive: it explains a phenomenon by appealing to relations between the parts of a system – in this case, interactions that occur between agents in a network.[32] But it does not

---

[31] Here there is a relationship both to "Earman's principle", as noted in footnote 29, and also to Butterfield (2011), who argues that in cases where one takes an unrealistic infinite limit, one should expect to see the qualitative behavior that arises in the limit appearing already on the way to the limit.

[32] Of course, one might consider stronger senses in which an explanation could be reductive. For instance, one might require that a reductive explanation gives us information about the details concerning the behavior of the micro-constituents of the system, or that a reductive explanation elucidate why the microscopic details are causally relevant for the

follow that the model supposes an *atomistic* conception of the economy, i.e. it does not determine the law governing the behavior of any arbitrary agent. Given some behavioral assumptions, it does constrain the kinds of structures they might reside in. In this case: hierarchical structures that (sometimes) exhibit discrete scale invariance. This does not require any particular arrangement of individuals because those particular details are in some sense irrelevant; what does matter are these structural details.

Likewise, the model is representational in the sense that its success depends on the fact that it represents certain stylized facts about market participants: they influence one another, at least sometimes, by imitation, and their interactions are hierarchical, in the sense that some traders are able to influence larger groups than other traders. Of course, this is far from a complete or accurate representation of market participants. But if actual market participants do not bear relations to one another that are adequately represented by a network with these features – or if markets are not discretely invariant across at least some scales – then the JLS model would fail to support the causal explanation we have described here. And so, it seems that the success of the explanation *does* depend on the representational accuracy of the model, at least with regard to these particular features.

This weak sense of being "representational" indicates that the JLS model may (also) be understood as an example of what Weisberg (2007, 2012) calls "minimalist models": "[A] minimalist model contains only those factors that make a difference to the occurrence and essential character of the phenomenon in question" (Weisberg 2007, p. 642). It also invokes Strevens' (2008) account of idealized models: "the content of an idealized model, then, can be divided into two parts. The first part contains the difference- makers for the explanatory target… The second part is all idealization; its overt claims are false but its role is to point to parts of the actual world that do not make a difference to the explanatory target" (318). Strevens, too, argues that this sort of idealization is compatible with causal explanation.

Of course, Batterman and Rice's *minimal* models and Weisberg's *minimalist* models are supposed to be fundamentally different; worse, those philosophers who have mistaken minimal models for minimalist models have "almost universally misunderstood" the explanatory structure of these models (Batterman and Rice 2014, 349). And yet, it would seem that the JLS model is an example of both. How could this be?

The tension can be resolved if one distinguishes between, on the one hand, features of a model – what sorts of idealizations it involves; in what senses, if any, it is representational; what sorts of mathematical relationships and methods it relies on – from the sorts of explanations one can give by appealing to the model – i.e., the why questions one is able to

---

phenomena under study. One might even insist that an explanation is reductive only if it appeals to fundamental physics – in which case, *no* explanation in the social sciences, and few in biology, chemistry, or even physics could ever be reductive. As we hope is clear from the text, we have in mind a weaker sense of an explanation being "reductive"; it is not essential to our purposes that this sense of reductive contravene Batterman and Rice. We are grateful to an anonymous referee for pushing us on this point.

answer (Fraassen 1980).[33] Batterman and Rice *define* minimal models as models used to give certain sorts of explanations involving universality classes. Since the JLS model can be used to explain why market crashes form a universality class (in the broad sense), the JLS model counts as a minimal model. These explanations, they argue, are neither causal nor reductive, and their success does not depend on the accuracy with which the models represent target systems; using the JLS model to explain the universal behaviors associated with crashes (namely, LPPLs, discrete scale invariance, etc.) is presumably also non-causal, at least insofar as Batterman and Rice's arguments are convincing.[34]

But the fact that the JLS model can be used for this sort of explanation does not bear on whether one can also use it to provide *other* explanations; nor does it bear on which explanations seem most salient in the context in which the JLS model was developed.[35]

In other words, we claim that the JLS model may be used to answer the question, "Why do markets generically exhibit volatility clustering, log-periodic oscillations, etc. near market crashes, even though market conditions otherwise vary dramatically?" To do so, one uses RG methods to show that a large variety of different networks exhibit discrete scale invariance and satisfy LPPLs near transitions points. In answering this question, we give the sort of explanation that Batterman and Rice are pointing to, and it is for this reason that the JLS model is a minimal model.

But we claim that we can *also* use the JLS model to answer the question, "Why do stock markets crash?", where this question is understood to be about the causes of crashes. And in this case, the answer is: because hierarchical networks can spontaneously evolve into states featuring discrete scale invariance, and scale invariance of any sort allows vanishingly small perturbations to cascade across scales.[36] It is in answering this question that the Woodwardian conditional described above is crucially invoked. And it is in answering this question that the minimalist representational features of the JLS model matter.

---

[33] This point mirrors one made by O'Connor and Weatherall (2016): there are many different purposes for which models may be constructed, and to which they may be put. This includes different explanatory purposes, and so one should be cautious about attempts to classify or taxonomize models on the basis of how they may be used to explain.

[34] We tend to think that they *are* convincing, or at least, we agree that explanations of universality of the sort Batterman and Rice discuss are non-causal. (See also Reutlinger 2014 for a different argument concerning why these explanations are non-causal.)

[35] We should emphasize that we do not take the claim that different questions call for different kinds of explanation to be in tension with Batterman and Rice's view. Our point, rather, is to resolve the apparent tension between our arguments and Batterman and Rice's view by distinguishing the why questions at issue. We are grateful to an anonymous referee for encouraging us to clarify this.

[36] Note that there is another interpretation of "Why do stock markets crash?" that does not demand a causal explanation, but rather another minimal model explanation: namely, "Why do markets fall into a universality class of systems that exhibit crashes, as opposed to tamer sorts of transitions?" Of course, this is a legitimate explanatory demand, and the answer, invoking the JLS model, would look more like the answer to the first question than the second. The difference between these two understandings of the question "Why do stock market crash?" invokes van Fraassen's (1980) analysis of the logic of why questions.

There are several points to emphasize here. The first is just to clarify our argument, lest our claims above be misconstrued: As should now be clear, when we argued above that the JLS model provides a causal explanation, we did *not* mean to imply that the explanation one can give for why market crashes form a universality class is a causal explanation (*contra* Batterman and Rice), nor (*ipso facto*) that *all* explanations are causal.[37] The point is rather that the JLS model, despite having the characteristic features of a minimal model, may nonetheless be used to give causal explanations (in addition to minimal model explanations). And pulling apart these different explanatory tasks requires careful attention to precisely what question one is trying to answer.

A second point to emphasize is that, even though the why questions described above are distinct, there is a subtle interplay between them. It is precisely *because* the JLS model can be used to explain why market crashes form a universality class in the relevant sense that it can (also) be used to provide a certain kind of causal explanation of market crashes, since it is the relationship picked out by this universality class, between discrete scale invariance and LPPLs near transition points, that makes true the conditional that forms the basis of the causal explanation. More, for precisely the same reason, the infinite idealization in the JLS model is essential precisely because it helps one identify the common mechanism underlying the phenomenon of interest – and thus, it is the infinite idealization that permits the causal explanation. Conversely, it is precisely because the relationship encoded by the Woodwardian condition holds that market crashes fall in a universality class (in the broad sense) in the first place.

This situation raises a question. If the JLS model can be both a minimal model and also a minimalist model, can we understand the other models that Batterman and Rice discuss, including models of critical phase transitions, as *also* providing interventionist causal explanations (in addition to minimal model explanations)? In a sense, the answer must be "yes", at least if what we argue above is correct. For instance, in the phase transition case, one can use the Ising model to answer the question, "Why do critical phase transitions occur?", construed causally, by showing that the Ising model, and a wide range of other models in its universality class, can evolve into states that are (approximately) scale invariant, and thus vanishingly small perturbations can cascade across scales. This explanation is causal in just the same sense that the corresponding explanation invoking the JLS model is. Once again, there is a subtle interplay between this explanation and the minimal model explanation using the same model, since the fact that real systems are in the same universality class as the Ising model is precisely what isolates scale invariance as the difference-maker (or, perhaps better, the manipulable variable).

---

Explanatory demands, van Fraassen convincingly argues, involve, in addition to the *explinandum*, both a *contrast class* and a *relevance relation*.

[37] For other examples of explanations that seem to be even more clearly *non*-causal, see Weatherall (2011, 2017).

All that said, there is still a difference between the JLS model and critical phase transitions in this regard. It concerns which explanatory demands seem most salient. As we noted above, one of the most striking features of critical phase transitions is the fact that many different systems have the same critical exponents. The salient issue is not to explain why transitions occur at all, but rather to explain why transitions in different systems are so similar. Of course, this does not prohibit one from asking the other question; it is just a matter of emphasis. (Besides, background theory, such as mean field theory, seems to explain this well, without explaining universality.) In the case of financial markets, the situation seems to be reversed: there, one wants to explain why (endogenous) market crashes occur *at all*, particularly given that crashes are often taken to be in tension with the EMH and other standard market modeling assumptions. And for this reason, it is the causal explanation using the JLS model that seems to be the salient one.

## 6. Policy Implications

We argued above, particularly in section 4.2, that the sense in which we take the JLS model to provide a causal explanation is interventionist: it depends on identifying a potential conditional relationship, the antecedent of which can be understood as a variable that can be manipulated, at least in principle. Moreover, the JLS model provides an observable signal of when that antecedent obtains. But having identified such a variable means that we have also identified a potential target for policy intervention. If we accept the JLS model, how might a regulatory agency intervene to prevent crashes? The answer is to disrupt the network structure on which traders reside.

How might one do this? One possibility would be through structural changes. Hierarchical networks have interesting dynamical properties because their inhabitants tend to cluster together and thus disseminate risk in particular ways.

> …hierarchical networks are resilient to peripheral crises, but very fragile in the face of crises in the center. In these systems, the risk of contagion falls as the system integrates around the center. (Oatley et al. 2013, p. 135)

Thus, one possible intervention would be to try to identify regions of the network that are peripheral, and try to introduce further connections – i.e., increase integration – between them, as this can make hierarchical networks more resilient to contagion.

It is not clear that this sort of proposal could serve as a response to an impending crash, however. Another proposal that might be more effective in this regard is given by Holme et al (2002). They borrow from computer science to suggest that sometimes the performance of a system can be improved by selectively deleting vertices and edges in a network (i.e. the relationships between nodes/agents):

> If one wants to protect the network by guarding or by a temporary isolation of some vertices (edges), the most important vertices (edges), breaking of which makes the whole network malfunctioning, should be identified. (1)

Here the suggestion would be to identify, in advance, particular relationships – say, relationships between major banks, or within banks – and intervene on them when LPPLs appear in market data, perhaps by blocking information from being exchanged between particular actors.

The JLS model can also be used as a diagnostic tool for evaluating current regulatory tools. For instance, one type of intervention that is actually used as a financial regulatory tool is the "trading curb". A trading curb works by temporarily halting activity if a very large, sudden drop occurs in the stock market. For instance, the New York Stock Exchange (NYSE) currently has in place several "circuit breakers," which kick in depending on how much the Dow Jones Industrial Average (DJIA) has moved within a short period of time, with longer time-out periods for larger sudden drops.[38]

> [T]he circuit-breaker halt for a Level 1 (7%) or Level 2 (13%) decline occurring after 9:30 a.m. Eastern and up to and including 3:25 p.m. Eastern, or in the case of an early scheduled close, 12:25 p.m. Eastern, would result in a trading halt in all stocks for 15 minutes. If the market declined by 20%, triggering a Level 3 circuit-breaker, at any time, trading would be halted for the remainder of the day. ("NYSE: NYSE Trading Information" 2016)

Circuit breakers may also be assigned to a particular stock, rather than to the market as a whole. For instance, "limit up, limit down" measures employed in some markets prevent a stock from being traded outside a certain price band for a few minutes (Pisani 2013). For instance, a 5% movement within five minutes (e.g. say a stock drops to $5 at that time) would mean that for 15 minutes, it would not be allowed to trade for less than $5.[39]

One motivation behind trading curbs is that in the period during the halt, investors will "calm down," i.e. behave more rationally rather than contributing further to a bubble of irrational exuberance (or pessimism). Unfortunately, some studies indicate that curbs can actually *encourage* such behavior, especially if agents know what the trading curbs are and whether the relevant limits are being approached (Goldstein and Kavajecz 2004). The JLS model provides some insight into why this might be. In particular, if stock markets crash because of long-range correlations between traders, then a trading curb merely slows down trading, without disrupting the underlying network state that causes the crash. Worse, the trading curb itself can serve as a coordinating signal to the entire network that the market is in a precarious state, in a way that actually *increases* correlations.

## 7. Conclusion

In the foregoing, we have argued that the JLS model provides a compelling causal explanation of market crashes, with potential predictive power. The model is consistent with mainstream models in financial economics, but clearly goes beyond them – and does so by exploiting an analogy with physics. As noted in the introduction, we take this as a proof of concept: econophysics at least has the capacity to contribute to our understanding of economic phenomena, even while remaining within the general realm of mainstream economic thought.

---

[38] Other exchanges, e.g. the Chicago mercantile exchange, have similar measures in place.

[39] This is actually a refinement of an old circuit breaker that would halt that stock's trade entirely for 5 minutes, but it caused too much administrative trouble to be usable.

We have also used the JLS model to explore how idealized models may be used to explain. We argue that the JLS model may be understood as *both* a minimal model and a minimalist model, and that the apparent tension between these accounts dissolves once one recognizes the different explanatory demands that a single model may be used to answer. The JLS model offers a causal explanation of why markets crash: namely, they crash because markets can evolve into states that are approximately discretely scale invariant, with long correlation lengths, such that small perturbations can have outsized effects. But this is not the only explanation one can give using the JLS model; one can also explain why crashes generically exhibit certain features, such as volatility clustering, by showing that crashes lie in a universality class, in the generalized sense described in the paper. That the same model may be used to offer two different explanations – one causal, and one, presumably, non-causal – points to the importance of separating questions concerning the explanatory purposes to which a model can be put from attempts to classify or characterize models themselves.

**Acknowledgments**

This paper is partially based upon work supported by the National Science Foundation under Grant No. 1328172. Previous versions of this work have been presented at the at a workshop on the Physics of Society and a conference on Infinite Idealizations, both hosted by the Munich Center for Mathematical Philosophy; we are grateful to the audiences and organizers for helpful feedback. We are also grateful to Didier Sornette for helpful discussions concerning his work and for detailed feedback on a previous draft of the paper, to Alexander Reutlinger for detailed comments on an earlier draft, and to two anonymous referees for their helpful comments.

Cowles, Alfred. 1933. "Can Stock Market Forecasters Forecast?" *Econometrica* 1 (3): 309. doi:10.2307/1907042.
Craver, Carl F. 2006. "When Mechanistic Models Explain." *Synthese* 153 (3): 355–76. doi:10.1007/s11229-006-9097-x.
Earman, John. 2004. "Curie's Principle and Spontaneous Symmetry Breaking." *International Studies in the Philosophy of Science* 18 (2–3): 173–98. doi:10.1080/0269859042000311299.
Fama, Eugene F. 1965. "The Behavior of Stock-Market Prices." *The Journal of Business* 38 (1): 34–105.
Fraassen, Bas C. Van. 1980. *The Scientific Image*. Clarendon Press.
Franklin, Alexander. 2017. "On the Renormalisation Group Explanation of Universality." *Philosophy of Science*. http://philsci-archive.pitt.edu/12654/.
Gallegati, Mauro, Steve Keen, Thomas Lux, and Paul Ormerod. 2006. "Worrying Trends in Econophysics." *Physica A: Statistical Mechanics and Its Applications* 370 (1): 1–6. doi:10.1016/j.physa.2006.04.029.
Gluzman, S., and D. Sornette. 2002. "Log-Periodic Route to Fractal Functions." *Physical Review E* 65 (3): 036142. doi:10.1103/PhysRevE.65.036142.
Goldenfeld, Nigel. 1992. *Lectures on Phase Transitions and the Renormalization Group*. Westview Press.
Goldstein, Michael A., and Kenneth A. Kavajecz. 2004. "Trading Strategies during Circuit Breakers and Extreme Market Movements." *Journal of Financial Markets* 7 (3): 301–33. doi:10.1016/j.finmar.2003.11.003.
Graf v. Bothmer, Hans-Christian, and Christian Meister. 2003. "Predicting Critical Crashes? A New Restriction for the Free Variables." *Physica A: Statistical Mechanics and Its Applications* 320 (March): 539–47. doi:10.1016/S0378-4371(02)01535-2.
Griffiths, Robert B., and Miron Kaufman. 1982. "Spin Systems on Hierarchical Lattices. Introduction and Thermodynamic Limit." *Physical Review B* 26 (9): 5022.
Holme, Petter, Beom Jun Kim, Chang No Yoon, and Seung Kee Han. 2002. "Attack Vulnerability of Complex Networks." *Physical Review E* 65 (5): 056109. doi:10.1103/PhysRevE.65.056109.
Ilinski, Kirill. 1999. "Critical Crashes?" *International Journal of Modern Physics C* 10 (04): 741–46. doi:10.1142/S0129183199000553.
Johansen, Anders, Olivier Ledoit, and Didier Sornette. 2000. "Crashes as Critical Points." *International Journal of Theoretical and Applied Finance* 03 (02): 219–55. doi:10.1142/S0219024900000115.
Johansen, Anders, Didier Sornette, and Olivier Ledoit. 1999. "Predicting Financial Crashes Using Discrete Scale Invariance." *Journal of Risk* 1 (4): 5–32. doi:10.21314/JOR.1999.01.4.
Joshi, Mark S. 2008. *The Concepts and Practice of Mathematical Finance*. 2nd edition. Cambridge ; New York: Cambridge University Press.
Kadanoff, Leo P. 2000. *Statistical Physics: Statics, Dynamics and Renormalization*. World Scientific.
Kaplan, David Michael. 2011. "Explanation and Description in Computational Neuroscience." *Synthese* 183 (3): 339. doi:10.1007/s11229-011-9970-0.
Kaplan, David Michael, and Carl F. Craver. 2011. "The Explanatory Force of Dynamical and Mathematical Models in Neuroscience: A Mechanistic Perspective*." *Philosophy of Science* 78 (4): 601–27. doi:10.1086/661755.
29